February 3, 1999
	Editors
	Physical Review D
	
	Dear Editors,
	               We are submitting the paper titled
	"Annual Modulation Signature for the Direct Detection
	of Milky Way Wimps and Supergravity Models" for publication
	in Physical Review D. The tex file of the paper is included
	below. ps file of 5 figures will be sent in separate e-mails.
	
	Sincerely ,
	
	Richard Arnowitt   Pran Nath
	               
\documentstyle[osa,manuscript,graphicx]{revtex}

\begin{document}

\def\lsim{\ ^<\llap{$_\sim$}\ }
\def\gsim{\ ^>\llap{$_\sim$}\ }
\def\r2{\sqrt 2}
\def\beq{\begin{equation}}
\def\eeq{\end{equation}}
\def\beqn{\begin{eqnarray}}
\def\eeqn{\end{eqnarray}}
\def\rmuu{\gamma^{\mu}}
\def\rmud{\gamma_{\mu}}
\def\PL{{1-\gamma_5\over 2}}
\def\PR{{1+\gamma_5\over 2}}
\def\sinW2{\sin^2\theta_W}
\def\AEM{\alpha_{EM}}
\def\mul{M_{\tilde{u} L}^2}
\def\mur{M_{\tilde{u} R}^2}
\def\mdl{M_{\tilde{d} L}^2}
\def\mdr{M_{\tilde{d} R}^2}
\def\mz2{M_{z}^2}
\def\c2b{\cos 2\beta}
\def\au{A_u}
\def\ad{A_d}
\def\cob{\cot \beta}
\def\v#1{v_#1}
\def\tb{\tan\beta}
\def\epem{$e^+e^-$}
\def\KK{$K^0$-$\bar{K^0}$}
\def\wi{\omega_i}
\def\xj{\chi_j}
\def\Wmu{W_\mu}
\def\Wnu{W_\nu}
\def\m#1{{\tilde m}_#1}
\def\mH{m_H}
\def\mw#1{{\tilde m}_{\omega #1}}
\def\mx#1{{\tilde m}_{\chi^{0}_#1}}
\def\mc#1{{\tilde m}_{\chi^{+}_#1}}
\def\mwi{{\tilde m}_{\omega i}}
\def\mxi{{\tilde m}_{\chi^{0}_i}}
\def\mci{{\tilde m}_{\chi^{+}_i}}
\def\mz{M_z}
\def\sw{\sin\theta_W}
\def\cw{\cos\theta_W}
\def\cb{\cos\beta}
\def\sb{\sin\beta}
\def\rwi{r_{\omega i}}
\def\rxj{r_{\chi j}}
\def\rfp{r_f'}
\def\Kik{K_{ik}}
\def\Fq2{F_{2}(q^2)}
%

\begin{center}{\Large \bf  
 Annual Modulation Signature for the 
Direct Detection of Milky Way wimps and Supergravity Models
 } \\
\vskip.25in
{R. Arnowitt$^a$ and  Pran Nath$^b$  }

{\it
a Center for Theoretical Physics, Texas A \& M University, \\
College Station, TX 77843, USA\\
b Department of Physics, Northeastern University, Boston, MA 02115, USA\\
}

\end{center}

\begin{abstract}                
An analysis is given of the annual modulation signal for the 
direct detection of relic neutralinos within the framework of 
supergravity unified models. It is shown that both the minimal 
and the non-minimal SUGRA models can generate  neutralino-proton 
cross-sections at the level compatible with the signals
 reported in the DAMA experiment at the Gran Sasso National
 Laboratory. Effects of proton stability on the analysis of the DAMA 
 data in the minimal and the non-minimal SUGRA models are also
discussed. 

\end{abstract}

\section{\normalsize \bf  INTRODUCTION}  
\hspace{.3in}Last year, the DAMA experiment examined the possibility of
the direct detection of Milky Way wimps using the annual modulation signal.
Based on a set up of approximately 100 kg of radiopure $NaI (T\ell)$
detectors in the Gran Sasso National Laboratory and 4549 kg-day of data,
they found an indication of such a signal\cite{bernabei1}. 
 More recently, an additional
14,962 kg-day of data has significantly strengthened the statistical
analysis favoring the presence of a yearly modulation 
signal\cite{bernabei2} with a wimp
mass and proton cross section of

\begin{equation}
M_w = (59^{+17}_{-14}) GeV; \hspace{.05in} \xi  \sigma_{w-p} =
(7.0^{+0.4}_{-1.2}) \times 10^{-6} pb
\end{equation}

\noindent where $\xi = \rho_w $/$\rho_0, \hspace{.03in} \rho_w$ is the local
Milky Way wimp mass density and $\rho_0 = 0.3
\hspace{.03in}GeVcm^{-3}$.

In the following, we will analyze the possible consequences of such a signal
within the framework of supergravity grand unification models with R-parity
invariance and gravity mediated supersymmetry (SUSY) 
breaking\cite{chams,applied}.  Such
models automatically predict the existance of cold dark matter (CDM) in the
universe, i.e. the relic lightest supersymmetric particle (LSP) remaining
from the Big Bang.  Further, over most of the SUSY parameter space, the LSP
is the lightest neutralino, $\tilde{\chi}^0_1$, and the calculated relic
density of $\tilde{\chi}^0_1$ is in accord with astronomical estimates over
a significant part of the parameter space.  For the relic density, we will
assume the range $0.05 \leq \Omega_{\chi^0_1} h^2 \leq 0.30$, where
$\Omega_{\chi^0_{1}}$ = $\rho_{{\chi}^0_1} / \rho_c$, where
$\rho_{\chi^0_1}$ is the mean relic density in the universe, $\rho_c = 3
H^2  /  8\pi G_N,   G_N$ = Newton's constant, and the Hubble constant H is
parameterized by $H = (100    km    sec^{-1} Mpc^{-1}) h$.

The annual modulation effect arises due to the motion of the Earth around
the Sun.  Thus, $\upsilon_E$, the velocity of the Earth relative to the
Galaxy  is $\upsilon_E = \upsilon_S + \upsilon_0    \cos   \gamma    \cos
\omega (t - t_0)$ where $\upsilon_S$ is the Sun's velocity relative to the
Galaxy $(\upsilon_S = 232 km / s)$,  $\upsilon_0$ is the Earth's orbital
velocity around the Sun $(\upsilon_0 = 30    km/s)$ and $\gamma$ is the
angle of inclination of the plane of the Earth's orbit relative to the
galactic plane $(\gamma\cong 60^o)$.  One has $\omega = 2 \pi/T$ ($T = 1$
year) and the maximum velocity occurs at day $t_0$ = 155.2 (June 2).  
The change in the
Earth's velocity relative to the incident wimps leads to a yearly modulation
of the scattering event rates of about 7\%.

The calculation of the $\tilde{\chi}^0_1 - p$ cross section proceeds as
follows\cite{jungman}.  One first calculates the relic density of $\tilde{\chi}^0_1$,
limiting the SUSY parameter space so that the above constraints on
$\Omega_{\chi^0_1}h^2$ are obeyed.  Within this constrained parameter space,
we then calculate the $\tilde{\chi}^0_1 - p$ cross section for incident halo
$\tilde{\chi}^0_1$ on the terrestial target.  In comparing the theoretical
$\sigma_{\chi^0_1-p}$ with the data, a number of uncertainties arise due
mainly to the lack of knowledge of input parameters.
We estimate an error of a factor of 2-3 in these calculations. In 
addition,
$\rho_{\chi^0_1}$ may vary from about $(0.2 - 0.7) GeV cm^{-3}$ (i.e., $0.7
\stackrel{<}{\sim} \xi \stackrel{<}{\sim} 2.3$) . 

Supergravity models have a wide range of applicability.  Thus, once one
phenomena begins to fix the SUSY parameters, it effects predictions in other
areas.  We will examine here the effects the DAMA data has on SUSY mass
spectrum predictions at future accelerator searches.  In addition, most
supergravity models predict the existence of proton decay.  While the
predictions  for proton decay are more model dependent than other parts of
the theory, there is a strong correlation in the SUSY parameter space
between DM detector event rates and the expected proton  
lifetime\cite{limits}, and we
will discuss this below.

\section{\normalsize \bf  SUPERGRAVITY MODELS}    

\hspace{.3in} We consider supergravity (SUGRA) models where supersymmetry is
broken in a hidden sector at the Planck scale $(M_P = 2.4 \times 10^{18}
GeV)$ by supergravity interactions and transmitted to the physical sector by
supergravity, giving rise to soft breaking terms\cite{chams}.  
If the hidden sector
interactions are generation independent, one obtains the simplest model,
mSUGRA, with universal soft breaking parameters at the GUT scale $M_G \simeq
2 \times 10^{16}  GeV$.  The renormalization group equations (RGE) then show
that the SUSY soft breaking at $M_G$ generates $SU(2) \times U(1)$ breaking
at the electroweak scale.  This model then depends on four parameters and
one sign (in addition to the parameters of the SM).  One can choose these
parameters to be the following:  $m_0$, the universal scalar soft breaking
mass at $M_G$; $m_{1/2}$, the universal gaugino mass at $M_G$ (or
alternately the gluino mass $m_{\tilde{g}} \cong (\alpha_3(M_Z) /  \alpha_G)
m_{1/2}$ where $\alpha_G \cong 1/24$ is the coupling constant at $M_G$);
$A_0$, the universal cubic soft breaking parameter (or alternately $A_t$,
the t-quark parameter at the electroweak scale); and $\tan \beta =
\hspace{.05in} <H_2>\hspace{-.05in}/\hspace{-0.05in}<H_1>$ where $<H_2>$
gives rise to the up quark masses and $<H_1>$ to the 
down quark and the lepton masses.
The RGE then determines $\mu^2$ (where $\mu$ is the Higgs mixing parameter
in the superpotential term $\mu H_1 H_2$), leaving the sign of $\mu$
arbitrary.

The mSUGRA model contains relatively few new parameters, and existing data
has begun to limit this parameter space.  Thus, the measured value of the
t-quark mass and the $b \rightarrow s + \gamma$ branching ratio eliminate
most of the $\mu < 0$ and $A_t < 0$ part of the parameter 
space\cite{bsgamma}. However, the supergravity formalism allows for 
non-universal soft breaking\cite{soni}. While the 
universality of the soft breaking mass $m_0$ guarantees the suppression of
flavor changing neutral currents (FCNC), the FCNC are not sensitive to
the Higgs mass or to the 
third generation non-universalities and string models can
allow for such non-universality.  
We parameterize this possibility by the
following Higgs soft breaking masses at  $M_G$,

\begin{equation}
m^2_{H_1} = m^0_2(1 + \delta_1) ; \hspace{.09in} m^2_{H_2} = m^2_0 (1 +
\delta_2)
\end{equation}

\noindent and the following third generation sfermion masses:

\begin{displaymath}
m^2_{q_L} = m^2_0 (1 + \delta_3);   \hspace{.09in} m^2_{u_R} = m^2_0 (1 +
\delta_4);   \hspace{.09in} m^2_{e_R} = m^2_0 (1 + \delta_5)
\end{displaymath}

\begin{equation}
m^2_{d_R} = m^2_0 (1 + \delta_6);  \hspace{.09in} m^2_{\ell_L} = m^2_0 (1 +
\delta_7)
\end{equation}

\noindent where $q_{L} = (t_L, b_L), \hspace{.03in} u_R = t_R, \ell_L =
(\nu_L, e_L)$, etc. $m_0$ is the universal soft breaking mass of the first
two generations, and $\delta_i$ represents the non-universal deviations.
For soft breaking occurring above $M_G$, and for GUT groups having an SU(5)
subgroup (e.g. SU(N), $N \geq 5$, SO(N), $N \geq 10$, $E_6$, etc.) with
matter embedded into 10 and $\bar{5}$ representations in the usual way, one
has $\delta_3 = \delta_4 = \delta_5$ and $\delta_6 = \delta_7$.  We assume
in the following that $\mid\delta_i\mid \leq 1$.

The non-universal corrections enter sensitively in $\mu^2$.  For $\tan
\beta \leq 25, \delta_5, \delta_6$ and $\delta_7$ make only small
contributions, and a closed form expression can be obtained for $\mu^2$ at
the electroweak scale\cite{nonuni}

\begin{displaymath}
\mu^2 = \frac{t^2}{t^2 - 1} [(\frac{1 - 3 D_0}{2} +  \frac{1}{t^2}) +
(\frac{1 - D_0}{2} (\delta_3 + \delta_4) - \frac{1+D_0}{2}   \delta_2 +
\frac{1}{t^2}\delta_1)] m^2_0
\end{displaymath}

\begin{equation}
+ \frac{t^2}{t^2 - 1}   [ \frac{1}{2} (1 - D_0)  \frac{A_R^2}{D_0} + C_{\mu}
m^2_{1/2}] - \frac{1}{2} M^2_Z + \frac{1}{22}     \frac{t^2 + 1}{t^2 - 1}
S_o (1 - \frac{\alpha_1(Q)}{\alpha_G})+ 1~loop ~terms
\end{equation}

\noindent where $t \equiv \tan   \beta$, $C_\mu = \frac{1}{2} D_0 (1 - D_0)
(H_3 / F)^2 + e - g / t^2$, and $D_0 \simeq 1-m^2_t / (200   \sin   \beta)^2$.
$D_0$ vanishes at the t-quark Landau pole (for $m_t = 175 GeV, D_0 \leq
0.23$) and $A_R = A_t - m_{1/2} (H_2 - H_3 / F)$ is the residue at the
Landau pole ($A_R \cong A_t  - 0.61 (\alpha_3 / \alpha_G) m_{1/2})$, i.e.
$A_0 = A_R / D_0 - (H_3 /F) m_{1/2}$.  $S_0 = Tr Y m^2$ where $Y$ is the
hypercharge and $m^2$ are the squark and Higgs masses at $M_G$ of Eqs.
(2,3).  The form factors $H_2, H_3, F, e, g$ are given in Ref.\cite{ibanez}.

$\mu^2$ enters importantly in the theoretical predictions of dark matter
event rates.  Thus, if $\mu^2$ is reduced, the predicted rates generally
increase, and if $\mu^2$ is increased, they go down.  From Eq.(4), we see
that the non-universal parameters thus can play an important role, as for
one set of signs (i.e., $\delta_{1,3,4} < 0, \hspace{.09in}  \delta_2 >0)
\hspace{.03in}  \mu^2$ will be reduced, and  the opposite set 
 will increase $\mu^2$.

\section{\normalsize \bf COMPARISON WITH DAMA DATA}

\hspace{.3in}We compare in this section, the DAMA data with the theoretical
expectations for SUGRA models. The analysis is carried out using the 
constraints arising from the radiative breaking of  the electro-weak
symmetry and the accurate method for the computation of the relic
density\cite{accurate}.

\noindent {\bf (i) mSUGRA model} 

In this model, all the $\delta_i$ are set to zero, and the theory is very
restricted.  Fig. 1 shows the experimental bound of the DAMA data  for $
\sigma_{\chi^0_1 - p}$ (dotted lines) vs. $m_{\chi^0_1}$ for the spin
independent $\chi^0_1 - p$ cross section, where we have combined the 
$95\% CL$ bound of DAMA with the uncertainty in the Milky Way density
$\rho_{\tilde\chi_1^0}$ ($0.7\leq \xi \leq 2.3$).  (The $NaI$ detector is sensitive
only to the spin independent interaction.)  The solid curves are the maximum
and minimum 
theoretical  cross section $\sigma_{\chi^0_1 - p}$
for $\tan \beta \leq
30$ (as one scans over the allowed part of the parameter space). 
In general, the maximum cross section occurs for the maximum values 
of $tan\beta$, so that the upper solid curve occurs for $tan\beta =30$.
The dashed curve corresponds to $tan\beta=20$ and the dot-dash curve 
to $tan\beta=10$.

The general behavior of the theoretical curves follow from an interplay
between the $\tilde{\chi}^0_1$ early universe annihilation cross section
(leading to current relic density) and the $\tilde{\chi}^0_1$ - quark cross
section in the terrestial DM detector\cite{events}.  
In the relic density analysis, there
are two regions in the neutralino annihilation cross section.  For
$m_{\chi^0_1} \stackrel{<}{\sim} 50-60$, annihilation can occur rapidly through
s-channel Z and $h$ poles, allowing (and sometimes requiring) $m_0$ to be
large (e.g., $m_0 = (500 - 1000)  GeV$) so that the lower bound,
$\Omega_{\chi^0_1} h^2 > 0.05$, not be violated.  For
$m_{\chi^0_1}\stackrel{>}{\sim} 50-60 GeV$, the t-channel annihilation through
sfermion poles dominates (as one now has $2m_{\chi^0_1} > m_h, \hspace{.03in}
M_Z$) and one requires $m_0$ to be small (e.g., $m_0 \stackrel{<}{\sim} 150
GeV$) in order that the sfermions are sufficiently light so that there is
sufficient annihilation to ensure that the upper bound $\Omega_{\chi^0_1}
h^2 < 0.3$ not be violated. 
Thus as $m_{\tilde\chi_1^0}$ is increased past 50 GeV, $m_0$ decreases. 
 In contrast, the detector $\chi^0_1$-quark
cross section has a major contribution from the s-channel squark pole and so
increases as $m_0$ is reduced.  In addition, this cross section falls off
with increasing neutralino mass.

The above effects can be seen in Fig. 1.  The $\sigma_{\chi^0_1 - p}$ cross
section rises as $m_{\chi^0_1}$ increases and subsequently falls off 
with increasing $m_{\chi^0_1}$.  This
leads to a maximum theoretical cross section at $m_{\chi^0_1} \simeq 55
GeV$, which fortuitously is in the region where the DAMA detector is most
sensitive.  One can therefore accomodate the DAMA data for 
$tan\beta>8$. 

\noindent {\bf (ii) Models with Non-universal Soft Breaking }
Non-universal soft breaking can significantly 
effect the analysis,
since as discussed above, $\mu^2$ of Eq.(4) can be decreased or increased
depending on the signs of the $\delta_i$. Thus for the cases where
$\delta_1,\delta_2,\delta_3<0$ and $\delta_4>0$ one finds that
$\mu^2$ decreases, and  a smaller $\mu^2$  tends to enhance the
 event rates and the $\tilde\chi_1^0-p$ cross-section. We study 
 this case in Fig.2 where we consider  $\delta_1=-1= -
 \delta_2$ and $\delta_3=-1=\delta_4$. Here we find that for 
  $tan\beta\leq 30$ theoretical predictions
  often exceed the upper  limit of the $2\sigma$ 
   corridor of DAMA whereas for the mSUGRA case the upper limit of 
   theoretical predictions lie within the DAMA corridor (see Fig.1).
    Similarly the 
   theoretical predictions for the case $tan\beta\leq 10$ lie 
   significantly higher than the corresponding predictions for 
   the mSUGRA case. Infact in this case one can achieve consistency
   with DAMA for values of $tan\beta$ as low as 6.
   Further the allowed $\tilde\chi_1^0$ mass range consistent with the 
   DAMA modulation signal extends over a somewhat larger
  domain of $tan\beta$ relative to the mSUGRA case 
   as may be seen by comparison of Fig.1 and Fig.2.

\section{\normalsize \bf  CONSTRAINTS OF PROTON DECAY}

\hspace{.3in}Many SUGRA models which possess a neutralino dark matter
candidate, also give rise to proton decay.  While predictions of proton
decay are more model dependent than in other phenomena, it has been observed
that cosmological constraints strongly affect predictions for the proton
lifetime $\tau_p$, and correspondingly, experimental bounds on $\tau_p$
affect expected DM detector rates\cite{limits}.

We consider here GUT groups which contain an SU(5) subgroup with matter
embedded into the 10 and $\bar{5}$ of SU(5) in the usual way.  For these
models, proton decay proceeds mainly through the 
mode $p \rightarrow \bar{\nu} +K^+$, 
with a $u$ and $d$ quark converted into $\tilde{d}_i$ and
$\tilde{u}_i$ squarks ($i$ = generation index) by a t-channel chargino
($\chi^{\pm}_j$), and the squarks are then converted into a $\bar{\nu}$ and a
$\bar{s}$-quark by the B and L violating interactions of the superheavy
color triplet Higgsinos, $\tilde{H}_3$ 
and $\bar{\tilde{H}}_3$\cite{pdecay,acn}.  The proton
lifetime is then

\begin{equation}
\tau^{-1}_p = \Gamma (p \rightarrow \bar{\nu}K) = \sum_{i=e,\mu,\tau} \Gamma
(p \rightarrow \bar{\nu}_i K^+)
\end{equation}

\noindent In general, the second generation dominates and in order to get
the maximum lifetime, $\tau_{max}$, we assume that the third generation,
which enters with arbitrary phase and is about a 20\% effect, interferes
destructively with the second generation.  We will also limit the
$\tilde{H}_3$ mass to obey $M_{H_3} \leq 10 \hspace{.03in} M_G$, as larger
values of $M_{H_3}$ would be in the domain of strong gravitational effects
not being considered in SUGRA models.

The current Super
Kamiokande bound on $\tau_p$ for the $ p \rightarrow \bar{\nu}  K$ 
mode is\cite{takita}

\begin{equation}
\tau_p (p \rightarrow \bar{\nu} K) > 5.5 \times 10^{32} \hspace{.03in} yr;
\hspace{.09in} 90 \% \hspace{.03in} C.L.
\end{equation}

One may obtain a qualitative picture of the dependence of $\tau_p$ on the
SUSY parameters by noting that the dominant second generation contribution
is roughly scaled by\cite{acn}

\begin{equation}
\Gamma (p \rightarrow \bar{\nu} K) \sim \frac{1}{M^2_{H_3}}
 (\frac{m_{\chi^0_1} \tan \beta}{m^2_{\tilde{q}}})^2
\end{equation}

\noindent Thus large $m_{\chi^0_1}$ (i.e. large $m_{\tilde{g}})$, small
$m_{\tilde{q}}$ and large $\tan \hspace{.03in} \beta$ will destabilize the
proton.  This puts strong constraints on the allowed  region of the 
parameter
space when combined with the relic density constraint $0.05 \leq
\Omega_{\chi^0_1} h^2 \leq 0.30$. 
 The DAMA data puts additional
strong constraints on the mSUGRA models.

 In more general situations the Higgs triplet sector of the theory can be 
 more complicated and one can have many Higgs triplets which 
 mediate p decay. Thus as discussed in ref.\cite{multihiggs} 
 in the presence of many Higgs triplets one has
 a Higgs triplet mass term  $\bar H_iM_{ij}H_j$, and one can 
 make a redefinition
 of fields so that only the Higgs triplet $\bar H_1$ and $H_1$ couple to
 matter. The Higgs triplet couplings to matter then are of the 
 form   $\bar H_1J+\bar KH_1$,
 where $J$ and $\bar K$ are bilinear in  matter (i.e., quark and lepton 
 fields). If one integrates out the Higgs triplet fields,  the
 resulting baryon and lepton number violating dimension five 
  operator is given by\cite{multihiggs} 
$W_4=-\bar K(M^{-1})_{11}J$. 
Thus for a $2\times 2$ matrix M, a suppression by a factor of $\approx 3$
(e.g., by the choice $M_{11}=-2M_{22}=M$, $M_{12}=M_{21}=M$) in the
proton decay amplitude and a supression by a factor of $\approx 10$
 in the 
proton decay rate is easily obtained.  
 In this case the proton lifetime can be significantly 
 enhanced.  
 
 We discuss now the numerical analysis of $\sigma_{\tilde\chi_1^0-p}$ 
 with the inclusion of the proton lifetime constraint. 
 It turns out that with the imposition of the proton lifetime 
limits, one finds no points in the parameter space consistent with
DAMA data for the minimal SU(5) case. However,  a non-trivial 
region of the parameter space does exist for non-minimal SUGRA models
where one can  get consistency simultaneously with 
both the DAMA data on annual modulation and the proton lifetime limits. 
The non-minimalities  can arise in various forms. We have already
discussed 
non-minimalities in the soft SUSY breaking parameters where 
non-minimality implies going beyond mSUGRA to include non-universal
soft breaking.
However, since proton decay involves GUT physics one finds that 
there are other non-minimalities needed to generate acceptable physics
at low energy. Thus the minimal SU(5) 
model does not lead to satisfactory quark and lepton mass matrices. 
One may modify it, however, to generate, e.g. the Georgi-Jarlskog texture at
$M_G$, leading to reasonably correct quark and lepton masses. 
This requires additional GUT interactions which give 
 rise to an increase in $\tau_p$ of about a 
factor of 3-5\cite{texture}.  In
addition to this, there are other theoretical uncertainties in the
calculation of $\tau_p$, relating to input parameters, e.g. the three 
quark matrix element of the 
proton\cite{gavela}, and we 
estimate an additional uncertainty of a factor of 2-3. 
Thus a factor of 10 enhancement is possible for the p lifetime from
these factors.  
Further the structure of GUT physics is of course  largly
	unknown. In general as discussd above one can conceive
	of multi Higgs triplets 
	which participate in mediating p decay. 
	As discussed above a factor of ten enhancement
	in the proton lifetime can occur from this source without
	significant fine tuning. Including the enhancements discussed from
	effects of textures etc. one can generate a total enhancement of 
	a factor of $\sim 10^2$.
	
	An analysis of $\chi_1^0-p$ cross sections with proton
	decay constraints  including the above enhancement factors is given
	in Fig.3 for the case of universal soft SUSY breaking
	parameters. One finds that there is a small region
	of the parameter space  where
	both the DAMA and the proton lifetime constraints are
	satisfied. The allowed $\tilde\chi_1^0$ mass region is
	close to the peak of the DAMA experiment.   
	Next we carry out the analysis for models with
	non-universal soft SUSY breaking. The model which
	is favorable for both DAMA and proton stability is
	given in Fig.4. It corresponds to the case 
	$\delta_1=-0.5=-\delta_2$. The analysis shows 
	that with enhancements factors as discussed above
	one can achieve consistency with DAMA and proton
	lifetime limit over a broad range of $\tilde\chi_1^0$
	mass and within the range given by DAMA.  
	This model produces an interesting range of mass 
	spectra for the SUSY particles. Two of the supersymmetric
	particles which are likely candidates for discovery
	at colliders are the Higgs and the lightest chargino. 
	We exhibit the scatter plot of their masses in this
	model in Fig.5. Their mass ranges are such that they 
	should be accessible at TeV(33).

\section{CONCLUSION}
In this paper we have given an analysis of the annual modulation
signal observed by DAMA for the direct detection of dark matter
within the framework of supergravity models\cite{bottino}. We find that 
SUGRA models with and without non-universalities can easily
accomodate the range of $\tilde\chi_1^0-p$ cross-sections needed 
 to explain the observed signal. However, in SUGRA models with
 grand unification, proton lifetime limits impose additional constraints.
 Thus for the case of the minimal SU(5) mSUGRA model it appears
 not possible to explain the DAMA signal and simultaneously
 satisfy the current proton lifetime lower limits. However, 
 SUGRA models with textures and with more complex Higgs triplet 
 structure can allow  one to significantly enhance the p decay 
 lifetime. Similarly, SUGRA models with non-universalities also
 affect the proton lifetime as well as  $\tilde\chi_1^0-p$ cross-section. It is
 shown that within the context of these non-minimal SUGRA models
 one can achieve consistency with the signal and a satisfaction of the
 proton lifetime limits. The sparticle mass spectra is also 
 investigated. It is shown that the lightest particles
 in this model are the neutralino $\tilde\chi_1^0$, the light chargino
 $\tilde \chi_1^{\pm}$, and the light Higgs $h^0$ which should all be
 accessible at the TeV(33) and the LHC.\\

\noindent   
{\bf Acknowledgements} \\
 This research was supported in part by NSF grant PHY-9722090 and
PHY-96020274.\\ 

\noindent
{\bf Figure Captions}\\
Fig1. Plot of the maximum and minimum of $\tilde\chi_1^0-p$ cross-section 
vs neutralino mass for mSUGRA for various ranges of tanbeta when other 
parameters are allowed to vary over the ranges $m_0,m_{\tilde g}$
$\leq  1 TeV$ and $A_0$ is allowed to vary over the range consistent 
with electro-weak symmetry breaking. The various cases correspond to
$tan\beta\leq 30$ (solid), $tan\beta\leq 20$ (dashed),
and $tan\beta\leq 10$ (dot-dashed). The minimum dashed and dot-dashed
curves overlap the  minimum solid curve. The dotted curves give the
 experimental $95\% CL$ range implied by the DAMA annual modulation
  data combined with the uncertainty in $\xi$.
 
Fig2. Plot of the maximum and minimum of $\tilde\chi_1^0-p$ cross-section 
vs neutralino mass for the non-universal SUGRA model with 
$\delta_1=-1=-\delta_2$, $\delta_3=-1=\delta_4$ 
for various ranges of $tan\beta$ when other 
parameters are allowed to vary over the ranges $m_0,m_{\tilde g}$
$\leq  1 TeV$ and $A_0$ is allowed to vary over the range consistent 
with electro-weak symmetry breaking. The various cases correspond to
$tan\beta\leq 30$ (solid), $tan\beta\leq 10$ (dashed),
and $tan\beta\leq 8$ (dot-dashed). The minimum  dashed and dot-dashed
curves overlap the  minimum solid curve over most of the neutralino 
mass range. The horizontal curves are as in Fig.1. 

Fig3. Plot of the maximum and minimum of $\tilde\chi_1^0-p$ cross-section 
vs neutralino mass for mSUGRA exhibiting the proton lifetime constraint.
The maximum and minimum solid curves are  for $tan\beta \leq 30$
as in Fig.1. The region 
enclosed by the long dashed curve and the minimum solid line is
the region allowed under the proton life time constraint with an enhancement
factor of $10^2$ as discussed in the text. The region enclosed by the 
dot-dashed lines on the left and the region enclosed by the dot-dashed 
line and the minimum solid line to the right is the region allowed 
with proton life time constraint with an enhancement factor of 20.

Fig4. Plot of the maximum and minimum of $\tilde\chi_1^0-p$ cross-section 
vs neutralino mass for the non-universal SUGRA model with
$\delta_1=-0.5=-\delta_2$, and $\delta_3=0=\delta_4$ 
exhibiting the proton lifetime constraint.
The maximum and minimum solid curves are for the case when
$tan\beta\leq 30$ but without imposition of proton lifetime constraint.
The region enclosed by the long dashed curve and the minimum solid
line is the region allowed by the proton lifetime constraint with
an enhancement factor of $10^2$. The region enclosed by the 
dot-dashed lines is the region allowed by the  proton life time constraint
with an enhancement factor of 20. 

\noindent  
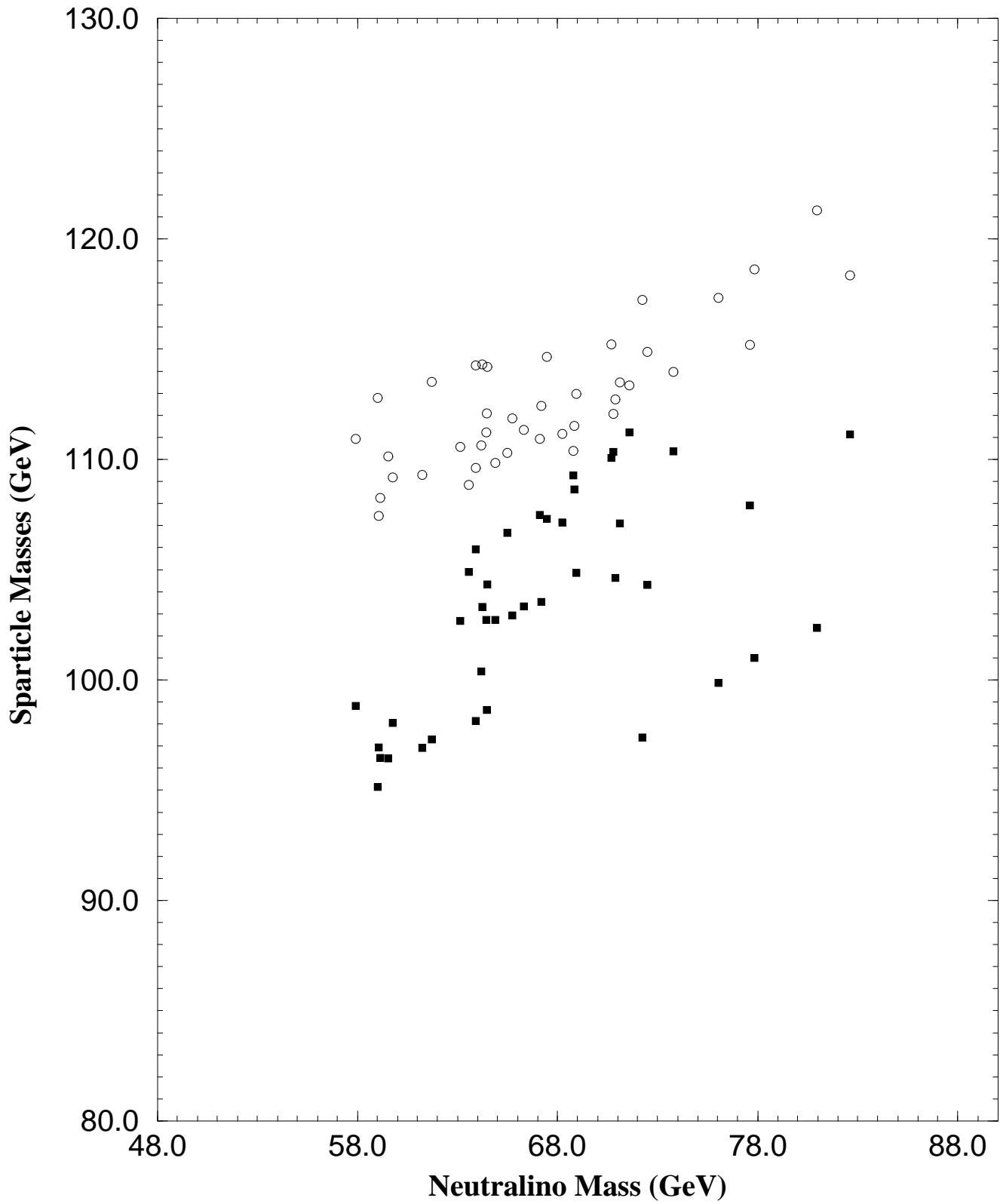
Fig5. The scatter plot of light chargino mass $\tilde W_1$ (filled 
squares) and the light Higgs $h^0$ (open circles) for the 
case $\delta_1=-0.5=-\delta_2$  for the
set of points that satisfy the DAMA range as shown in Fig.1 
  and the proton stability constraint corresponding to the
  area enclosed by the dashed curves in Fig4.

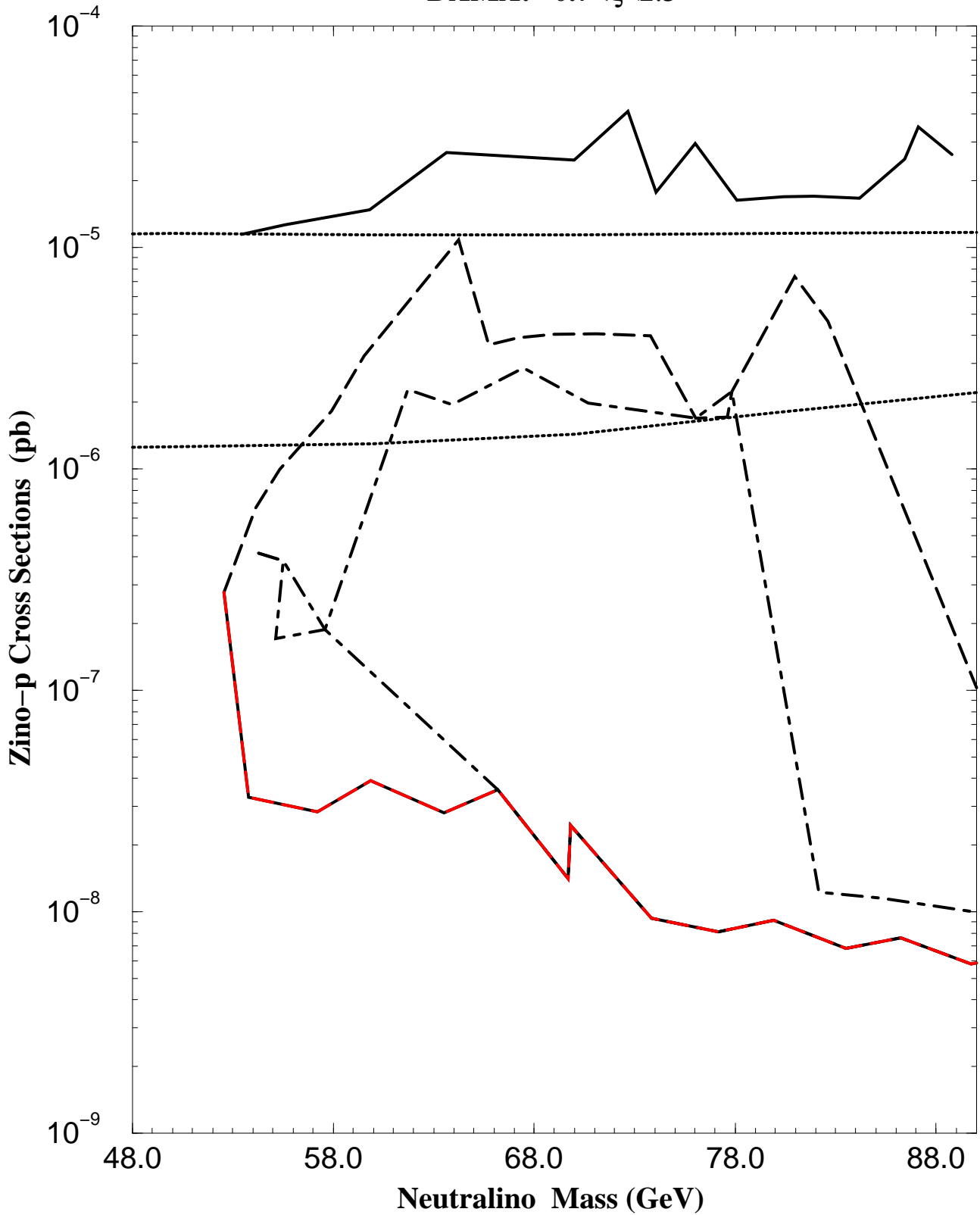
\end{document}